\documentclass[a4paper,fleqn]{cas-dc}

\usepackage[numbers]{natbib}

\usepackage{graphicx}
\graphicspath{ {./graphics/} }

\usepackage{threeparttable}

\def\tsc#1{\csdef{#1}{\textsc{\lowercase{#1}}\xspace}}
\tsc{WGM}
\tsc{QE}
\begin{document}
\let\WriteBookmarks\relax
\def\floatpagepagefraction{1}
\def\textpagefraction{.001}

% Short title
\shorttitle{Fallout}    

% Short author
\shortauthors{G. Bolton King et al}

% Main title of the paper
\title [mode = title]{Fallout: Distributed Systems Testing as a Service}  

% Title footnote mark
% eg: \tnotemark[1]
%\tnotemark[<tnote number>] 

% Title footnote 1.
% eg: \tnotetext[1]{Title footnote text}
%\tnotetext[<tnote number>]{<tnote text>} 

% First author
%
% Options: Use if required
% eg: \author[1,3]{Author Name}[type=editor,
%       style=chinese,
%       auid=000,
%       bioid=1,
%       prefix=Sir,
%       orcid=0000-0000-0000-0000,
%       facebook=<facebook id>,
%       twitter=<twitter id>,
%       linkedin=<linkedin id>,
%       gplus=<gplus id>]

\author[1]{Guy Bolton King}
\ead{guy@waftex.com}

% URL of the first author
%\ead[url]{<URL>}

% Credit authorship
% eg: \credit{Conceptualization of this study, Methodology, Software}
%\credit{Credits}

% Address/affiliation
\affiliation[1]{organization={DataStax Inc.}}
%           addressline={}, 
%           city={},
%          citysep={}, % Uncomment if no comma needed between city and postcode
%           postcode={}, 
%           state={},
%           country={}}

\author[1]{Sean McCarthy}
\ead{sean.mccarthy@datastax.com}

\author[1]{Pushkala Pattabhiraman}
\ead{pushkala.pattabhiraman@datastax.com}

\author[1]{Jake Luciani}
\ead{jake@datastax.com}

\author[1]{Matt Fleming}
\ead{matt@codeblueprint.co.uk}

% Footnote of the second author
%\fnmark[2]

% Email id of the second author

% URL of the second author
%\ead[url]{}

% Credit authorship
%\credit{Cred for matt}

% Address/affiliation
%\affiliation[<aff no>]{organization={},
%           addressline={}, 
%           city={},
%          citysep={}, % Uncomment if no comma needed between city and postcode
%           postcode={}, 
%           state={},
%           country={}}

% Corresponding author text
%\cortext[1]{Corresponding author}

% Footnote text
%\fntext[1]{}

% For a title note without a number/mark
%\nonumnote{}

% Here goes the abstract
\begin{abstract}
All modern distributed systems list performance and scalability as their core strengths. Given that
optimal performance requires carefully selecting configuration options, and typical cluster
sizes can range anywhere from 2 to 300 nodes, it is rare for any two clusters to be exactly
the same. Validating the behavior and performance of distributed systems in this large
configuration space is challenging without automation that stretches across the software
stack. In this paper we present Fallout, an open-source distributed systems testing
service that automatically provisions and configures distributed systems and clients,
supports running a variety of workloads and benchmarks, and generates performance reports
based on collected metrics for visual analysis. We have been running the Fallout service
internally at DataStax for over 5 years and have recently open sourced it to support our
work with Apache Cassandra, Pulsar, and other open source projects. We describe the
architecture of Fallout along with the evolution of its design and the lessons we learned
operating this service in a dynamic environment where teams work on different products and
favor different benchmarking tools.
\end{abstract}

% Use if graphical abstract is present
%\begin{graphicalabstract}
%\includegraphics{}
%\end{graphicalabstract}

% Research highlights
%\begin{highlights}
%\item 
%\item 
%\item 
%\end{highlights}

% Keywords
% Each keyword is seperated by \sep

\maketitle

% Main text
\section{Introduction}\label{}

Building databases and distributed systems with high performance requires thorough testing and
benchmarking. The earlier that performance testing can be done in the development
process, the cheaper issues are to fix \cite{Boehm}.

Software teams are now expected to use techniques such as CI/CD \cite{CONTINUOUS} to deliver
frequent releases to users. For many types of products, including distributed systems and databases,
users also expect the systems to be resilient, never lose data, and always achieve high performance.
Strong automated testing tools are required to reduce development time and deliver stable products.

Automating the testing of complex distributed systems requires tightly controlling every aspect of
the software: from operating system configurations to application-level tuning. Fallout evolved into
a full-stack orchestration system, enabling us to test and tweak all aspects of the distributed
system under test. Fallout is a service that deploys hardware resources, configures the operating
system and distributed application, runs a workload or benchmark on the cluster and gathers the
results for analysis. Through a rich YAML-based configuration, every aspect of the system and
application can be detailed and parameterized.

We use Fallout to run a mixture of manual and automated testing and Fallout executes around 200
tests every day. These tests have been used to verify the performance of new features and
optimizations, uncover functional and performance regressions before they have shipped to customers,
and reproduce issues that were discovered in the field. Recently, we have added support for chaos
testing too. Automated testing is driven by Jenkins which is the CI tool of choice for the majority
of our teams. The rest of this paper is organized as follows. In Section 2 we discuss our rationale
for building Fallout along with the existing tools at the time. In Section 3 we present a high-level
overview of the Fallout design and dive down into the details in Section 4. Section 5 illustrates
how Fallout test run results are displayed for users. Lessons learned, related work, and conclusions
are covered in Section 6, 7, and 8.

% Numbered list
% Use the style of numbering in square brackets.
% If nothing is used, default style will be taken.
%\begin{enumerate}[a)]
%\item 
%\item 
%\item 
%\end{enumerate}  

% Unnumbered list
%\begin{itemize}
%\item 
%\item 
%\item 
%\end{itemize}  

% Description list
%\begin{description}
%\item[]
%\item[] 
%\item[] 
%\end{description}  

% Figure
%\begin{figure}[<options>]
%	\centering
%		\includegraphics[<options>]{}
%	  \caption{}\label{fig1}
%\end{figure}

%\begin{table}[<options>]
%\caption{}\label{tbl1}
%\begin{tabular*}{\tblwidth}{@{}LL@{}}
%\toprule
%  &  \\ % Table header row
%\midrule
% & \\
% & \\
% & \\
% & \\
%\bottomrule
%\end{tabular*}
%\end{table}

% Uncomment and use as the case may be
%\begin{theorem} 
%\end{theorem}

% Uncomment and use as the case may be
%\begin{lemma} 
%\end{lemma}

%% The Appendices part is started with the command \appendix;
%% appendix sections are then done as normal sections
%% \appendix

\section{Background}\label{background}

Five years ago, we had a server-based performance testing and comparison tool named
cstar\_perf that could bootstrap Apache Cassandra onto an already provisioned cluster,
run a workload against it, and plot the performance results on a web page. The workload was composed
via a web UI and used cassandra-stress \cite{CASSANDRASTRESS} to generate load on the cluster. cstar\_perf gave us some
flexibility in that the Cassandra installation could be configured in a number of ways but it also
came with many limitations. The size of the cluster was fixed and could not be changed. The workload
consisted of a number of linear steps, each of which could invoke one of a small number of tools.
This gave us neither the modularity we needed to support diverse teams with different preferences
for benchmarks, tools, and workloads, nor the parallelism required to run multiple tests at once.

Fallout was conceptualized to address these limitations. There was a clear need to create a system
that could seamlessly stitch together a plethora of tools and systems built by internal teams so
they could be made to work together while remaining tool agnostic. It was also desired to provide
the ability to support any testing environment, be it public or private cloud. Since Fallout needed
to test distributed systems, it needed to support scenarios involving multiple server/client
clusters and a myriad topology configurations as well as tools that disrupt normal operation such as
throttling the network bandwidth and deleting cluster data. While cstar\_perf gave us the ability to
analyze performance for a single test run we also wanted the ability to generate better insights
into results by gathering artifacts from those clusters. To encourage adoption from a diverse set of
stakeholders, Fallout was required to be intuitive, simple, and self-documenting. The target user
group ranged from seasoned database engineers to non-developers. Hence, Fallout needed to use a
declarative language that was simple for non-developers to write tests in. The artifacts involved in
Fallout were required to be persisted and versioned for future reference. All of the test
configurations, results, and artifacts were to be stored in a single place so that everything could
be trivially shared within our organization.

In summary, Fallout addresses the following engineering challenges:

\begin{itemize}
\item Build a testing service that provides a single interface for multiple teams to run test and benchmarking tools
\item Use simple test configuration files to deploy tests into distributed systems that accurately reflect real-world configurations
\item Extract and preserve test run artifacts for later analysis
\item Ease of use for both developers and non-developers
\end{itemize}

The initial version of Fallout used Jepsen \cite{JEPSEN} as the workload execution tool. This was
largely a pragmatic choice since Jepsen was well-known in the original Fallout team and using it
avoided the need to reinvent the wheel by creating a brand new tool. Fallout extended Jepsen's
correctness testing features by creating operation logs during test runs and allowing pass/fail
checks to be run on test completion. Over time, Fallout has evolved into a more performance-focussed
service but still retains a couple of the original Jepsen concepts such as Checkers and operation
logs.

\section{Architecture}\label{architecture}

Fallout runs as a single service and exposes a REST API which is accessible via a Python client API
and command-line application, and a web UI which users can access using a web browser. Fallout
supports multiple concurrent users while enabling each user to store and execute tests
independently. Read-only access of test configurations is granted for other user's test
configurations which is especially handy when multiple engineers are working on the same test since
they can clone the test configuration and collaborate. The Python client API and command-line
application are used by Continuous Integration tools to submit tests to Fallout's job queue. Once a
job reaches the front of the queue and hardware resources become available, Fallout deploys and
configures the test's infrastructure (setup), runs the workload, then collects test artifacts and
tears down the infrastructure once the test is complete. Results are published to a central server
for analysis. Fallout maintains logs of all the operations involved in each step of a test. An
overview of Fallout's architecture is given in Figure \ref{architecture}.

% Figure
\begin{figure*}
\begin{center}
	\includegraphics[width=0.70\textwidth]{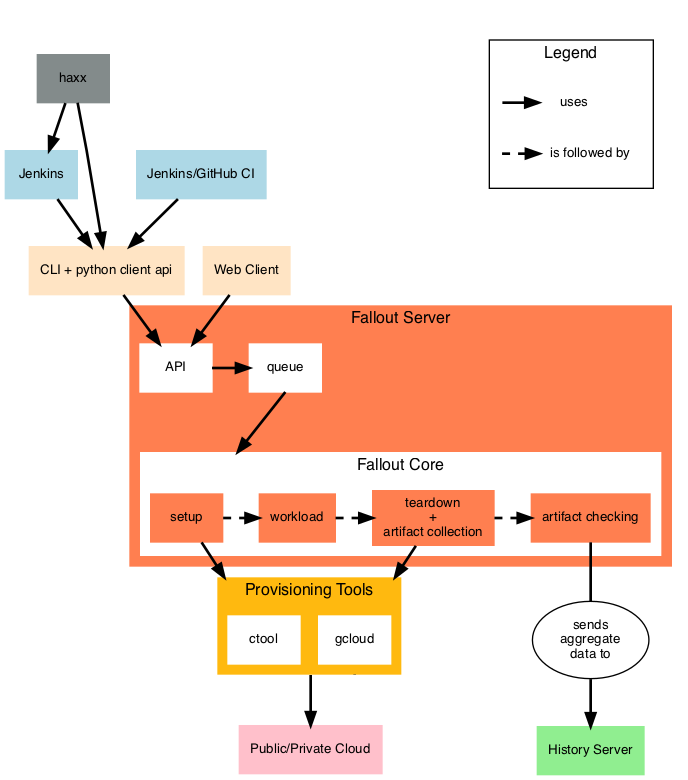}
	\caption{Fallout Architecture}\label{architecture}
\end{center}
\end{figure*}

\subsection{Cluster Deployment}

Test jobs are submitted to Fallout which internally schedules them based on the available hardware
resources in the infrastructure. To provision the cluster in DataStax's data center (private and
public), Fallout relies on a proprietary infrastructure tool, \emph{ctool}. The open-source version
of Fallout includes support for using Google Kubernetes Engine (GKE) to manage clusters.
\emph{ctool} is cloud provider agnostic and abstracts the provisioning and deployment steps of
Fallout tests so that users only need to specify high-level requirements such as cloud provider,
instance type and region in a YAML test config file. Fallout handles provisioning machines with GKE
using the gcloud tool \cite{GCLOUD} and includes logic for configuring resources that might be
required for the test. For example, Fallout will automatically add persistent storage to the
Kubernetes cluster so that test run artifacts can be downloaded from the cluster once the test
completes. Users can also specify custom manifests in their test config files which configure
cluster resources. Fallout monitors all logs from the cluster and can display them in real time via
the Fallout web UI. Once the test completes and the cluster is torn down, those logs are permanently
stored on the Fallout server for offline analysis.

Running performance tests against clusters requires applying workloads and benchmarks. Fallout also
handles provisioning and configuring client nodes that generate these workloads. Metrics and
statistics are gathered for all the client and server nodes via a dedicated observer instance that
is configured for the test run in exactly the same way as both client and server: via the test
config. In each test, the observer instance operates for the duration of the test run and allows
Fallout users to monitor metrics from the client and server in real time. Watching the live observer
node is frequently important when re-running a configuration that is known to exhibit performance
issues and the observer can be used to detect when a cluster has entered a bad state of performance.
At the end of the test, the observer metrics are archived and saved locally to the Fallout server
and available on the test run web page. This enables analysis after the test execution has
completed. Lastly, Fallout tears down the infrastructure after the test completes thereby returning
the allocated resources to the cloud.

\subsection{Application Installation, Configuration, and Execution}

The specific method used to install applications such as Apache Cassandra and Pulsar varies between
releases and engineers are often unaware of the differences. Fallout automatically handles
installation and system configuration no matter which version is specified for the test.
Installation involves extracting tarballs on each node and updating the \emph{cassandra.yaml} config
file to use the additional larger disks from the deployment phase -- Fallout also needs to handle
configuration of each individual node to work in the cluster. For instance, Apache Cassandra
requires the IP addresses of seed nodes in a cluster to be known and listed in every node's config
file.

Benchmarking tools including profilers and metrics collection agents are installed on the client
nodes by Fallout. Fallout supports a wide variety of tools though only a few of them are currently
available in the open source version. We plan on contributing more in the future. Each benchmark can
be configured using the same YAML interface and individual options contained in the config will be
specific to each benchmark. As Fallout has gained popularity, more and more benchmarks have been
added since it is common for different teams to favor different benchmarks. For example, YCSB is a
popular open source benchmark often used to compare relative performance of NoSQL database
management systems. The DataStax Stargate team use YCSB to benchmark Stargate’s Document API
performance for every release.

Fallout was designed to accommodate this heterogeneity while still providing the same interface to
users. This has an added benefit -- because the complexity of supporting multiple benchmarks is
primarily hidden inside of Fallout, external services that use Fallout can automatically work with
any benchmark, reducing the effort required to support new teams and new tools.

\subsection{Artifact Collection and Analysis}

To aid with post-run analysis, Fallout saves a range of logs and other artifacts locally on the
central server so that they can be inspected after the test run has finished. This is the most
common situation for analyzing metrics and other benchmark data collected as part of a manual test
run. For automated test analysis, Fallout will push the archived metrics to a central Grafana server
where other tools run further analysis on them, including Hunter, our statistical significance
detection tool that uses change point detection \cite{CPD}. Fallout uses
artifact checkers to inspect the logs for specific error or warning messages and allows the test run
to be marked as failed if any are present. Other artifact checkers are used to post-process files.
For example, the hdrtool artifact checker merges HDR files \cite{HDRHISTOGRAM} retrieved from multiple clients and
produces aggregated metrics.

Even when a performance regression is automatically detected by Hunter, engineers might need to look
at the metrics that were collected during the test run to understand the cause of the performance
issue. When a user needs to check the observer metrics they can simply download the archived
artifact from Fallout, extract it to their machine and use a docker image containing Grafana to
display the metrics.

\subsection{Integration with CI}

Automated testing with Fallout is primarily driven via Jenkins. Jenkins uses the Fallout API to
launch test runs whenever a pull request from GitHub is successfully built. We have configured
Jenkins so that it links directly to the Fallout test run for a given job (GitHub pull request).
Being able to navigate from the Jenkins job to the Fallout test run acts as a breadcrumb trail and
simplifies post-test run analysis.

We also run nightly and weekly performance tests that are scheduled outside of the GitHub PR-merge
workflow but still rely on Jenkins to call the Fallout REST APIs. Haxx is a git repository that acts
as a central location for storing Fallout test configs since Fallout itself does not provide any
kind of version control other than A) storing a read-only copy of the YAML file from previous test
runs and B) the most recent version. Haxx also provides templating for Fallout YAML files where
common configuration snippets, such as optimal Apache Cassandra configuration options, can be stored
in template files and reused across test configs. This allows us to significantly cut down on the
boiler plate code required to support a large number of tests where only the machine size, version
of Apache Cassandra, or benchmark config is different. Better still, templates allow users to take
advantage of known-good performance options which ensures that they do not waste their time analyzing
performance issues that were the results of poorly configured tests.

\section{Implementation}\label{implementation}

Since Fallout was originally created as a wrapper around Clojure, Fallout had to be written in
another JVM language to make development easier and Java was selected as the target language.
Despite Fallout development primarily being the responsibility of a very small team, Fallout has
benefited from a large number of contributors and since Java is widely used inside of DataStax the
choice of programming language is no doubt a contributing factor.

A similar desire to make the configuration interface as welcoming for users as possible led to the
decision to use YAML for the configuration files. YAML syntax is easy to learn for new users and
YAML syntax highlighting is readily available in IDEs and editors. Fallout's web UI provides a
built-in YAML editor with syntax checker for creating and modifying test configurations.

\subsection{Test Configuration Files}

Fallout test runs are driven by a single YAML configuration file that has a number of required
entries. Tests describe machines and services running on those machines. A node is a resource with
services running on it. An example of a node is a single Apache Cassandra node within a multi-node
cluster. NodeGroups are collections of nodes. An example of a NodeGroup is an Apache Cassandra
cluster. An ensemble is a set of NodeGroups with a specified role and test run configuration files
expose this concept to the user. The list of ensemble roles is:

\begin{itemize}
	\item Server: A distributed server or cluster such as Apache Cassandra
	\item Client: A benchmark or workload
	\item Observer: A monitoring server such as graphite
	\item Controller: An external controller such as Jepsen
\end{itemize}

\begin{figure*}
\begin{center}
	\includegraphics[width=0.7\textwidth]{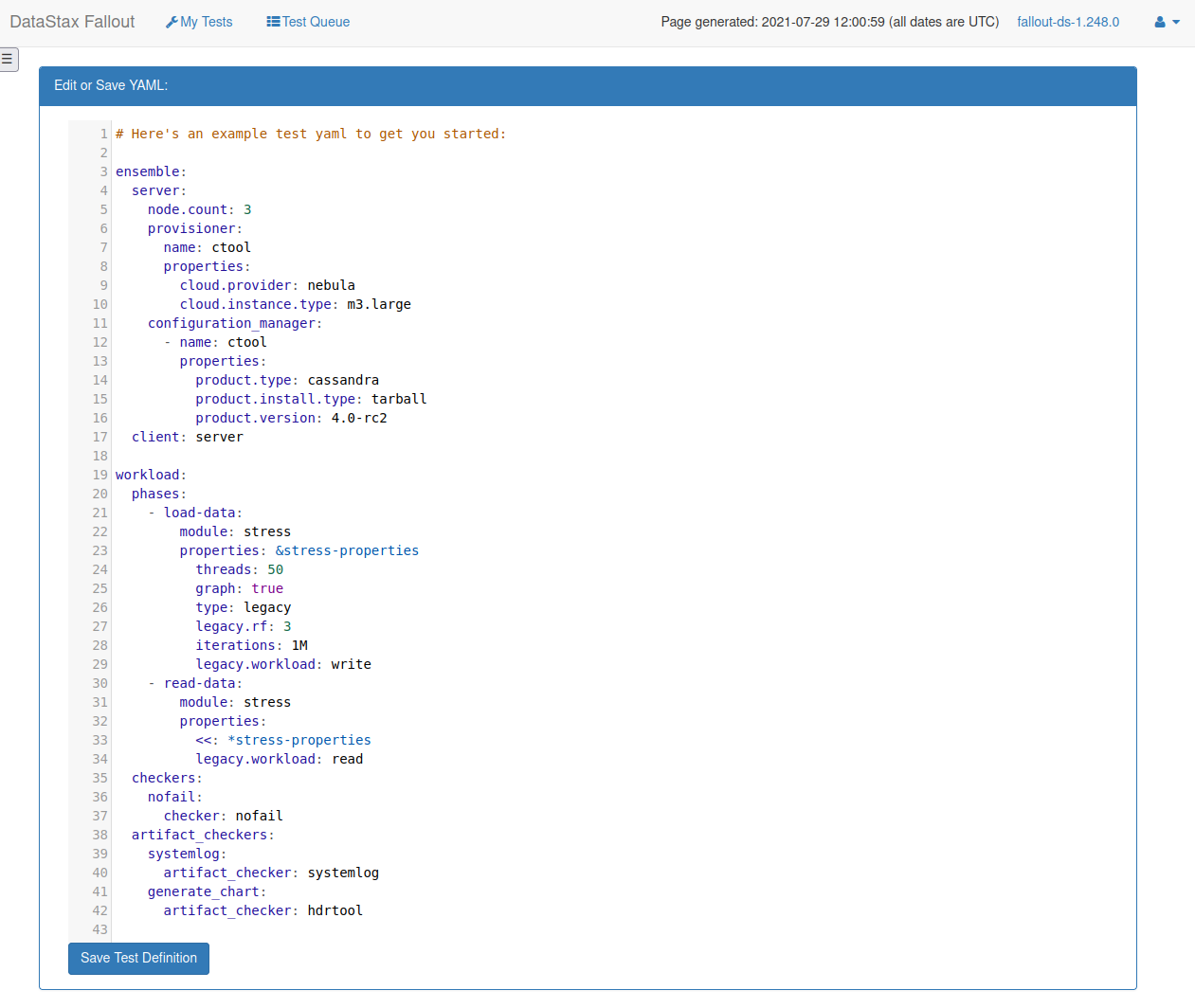}
	\caption{Example Fallout test configuration}\label{example}
\end{center}
\end{figure*}

Figure \ref{example} shows an example of a Fallout test configuration file.

% Figure

Workloads are built from one or more phases which are the basic unit of concurrency in Fallout. Each
phase can run one or more modules and specifying more than one module executes them in parallel.
Phases are always run sequentially and a phase will not start executing until the previous phase
completes.

\subsection{Test Provisioning Lifecycle}

Each NodeGroup in a test transitions through a number of states when the test executes. There are
three types of states: Unknown, Transitional, and Runlevel. Transitional states are entered when a
NodeGroup moves from one state to another. Runlevel states represent steady states where a NodeGroup
is not currently transitioning and are modelled on the UNIX runlevel concept -- NodeGroups progress
to higher levels where each level has more capabilities than the previous one. State transitions
perform provisioning and configuration actions on the NodeGroup and the current state of a NodeGroup
is used by Fallout to guarantee only legal transitions between states can occur. Using
the state machine, it is impossible for Fallout to configure a NodeGroup before it is provisioned. If
any errors are encountered during a transition, for example if Fallout fails to install the
distributed application, the NodeGroup will enter the FAILED state and the entire test run will
fail.

\begin{figure}
	\includegraphics[width=0.45\textwidth,height=12cm]{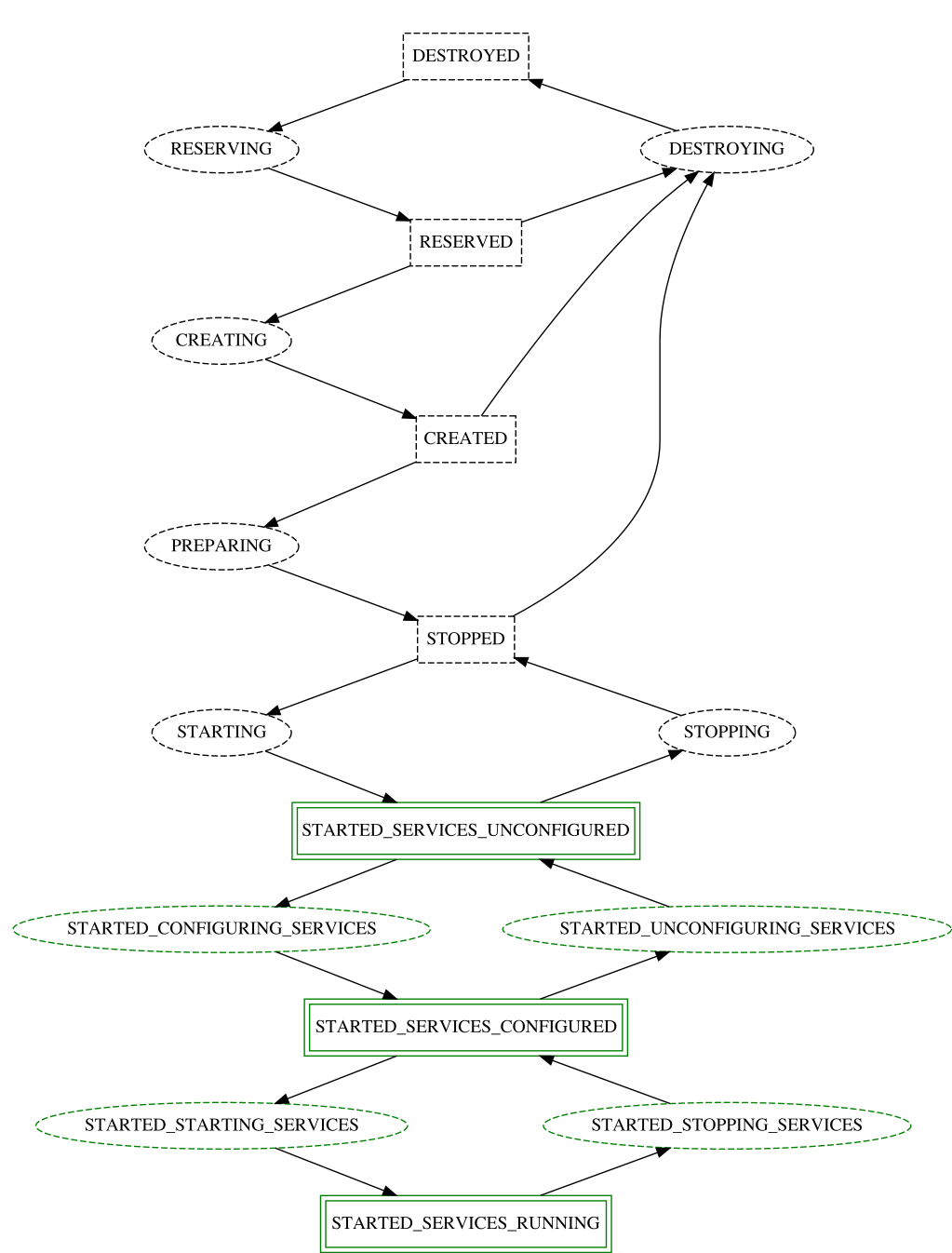}
	\caption{NodeGroup Transition Diagram}\label{nodegroup}
\end{figure}

A transition diagram is presented in Figure \ref{nodegroup}. The oval states on the left and right represent
Transitional states, and the rectangular states in the center represent runlevel states.

\subsection{Modules, Providers, and Configuration Managers}

Adding support for a new benchmark or tool to Fallout requires adding 3 new components to the
Fallout code base: a module, a provider, and a configuration manager. Providers allow access to a
service or tool via an API and these are invoked by the Fallout test harness to run commands on the
node. For example, the \emph{NoSqlBenchPodProvider} is responsible for executing the nosqlbench
\cite{NOSQLBENCH} benchmark on a Kubernetes pod. Providers can also have dependencies on other
Providers which makes it possible to express that a benchmark should only be available when running
on a Kubernetes cluster, for example. Fallout supports Chaos Mesh \cite{CHAOSMESH}, a tool for
running chaos experiments on a cluster, however since it is only available on Kubernetes Fallout
will refuse to deploy it into any environment that does not meet the Kubernetes Provider dependency.

Configuration Managers are responsible for configuring and unconfiguring software running on nodes
as well as starting and stopping services. Additionally, Configuration Managers register Providers
with nodes, making the associated services available to Modules in a test workload.

Finally, Modules are the user-facing component of benchmarks. Modules define the supported keywords
and parameters that can be passed to the benchmark via YAML configuration files. Since this
provides a layer of indirection between the test config and the benchmark itself, it is common for
only a subset of the parameters supported by the benchmark to be supported in Fallout, though if
users want maximum flexibility there is usually an \emph{args} parameter that passes through parameters
without any kind of filtering.

While Fallout supports a number of different benchmarks, one lesson we have learned is that users need
some kind of back-stop module that allows them to manually run benchmarks for which no support
currently exists. A bash module is provided to fill the gap where users need to run a simple script
or download a benchmark to a node and run it manually. Extended use of the bash module is frowned
upon because we have seen it lead to difficult to understand shell scripts that are copied between
test configs.

\subsection{Checkers and Artifact Checkers}

Once a test has completed, Fallout needs a way to validate that the system under test behaved
correctly for the duration of the test. Checkers are the component in Fallout responsible for
ensuring that no errors occurred during the test that might invalidate the results. This is
important for performance tests even though the checkers do not perform any kind of performance
analysis themselves -- any performance results from tests that fail basic checks are likely to be
invalid because the test was not run under real-world conditions. \emph{NoFailChecker} is an example of
a very basic checker that simply checks that none of the Fallout operations that ran during a test
failed. The history of operations is passed to checkers as an argument so that they can run arbitrary
checks against it. There is no limit to the number of checkers that can be included in a Fallout
test and a test will only pass if all checkers pass.

A related concept is the artifact checker which performs the same kind of validation process on
artifacts that are collected after the test run completes. A frequently used artifact checker for
Apache Cassandra tests is \emph{SystemLogChecker} which checks Cassandra's \emph{system.log} for the
presence of user-specified patterns such as log messages containing “ERROR” or “WARN”.

\subsection{Test Queue}

\begin{figure}
	\includegraphics[width=0.45\textwidth]{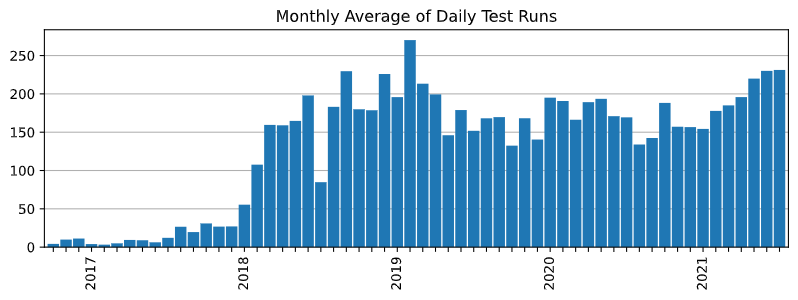}
	\caption{Average daily test runs by month}\label{testruns}
\end{figure}

When Fallout was first launched, test runs were executed as soon as they were submitted. As Fallout
grew in popularity, contention for VMs on our internal infrastructure resulted in tests failing. A
simple queueing mechanism was added to fix this that checked for VM availability before attempting
to submit a claim for resources. It has been tweaked over time to become more robust and fair.
For example, it now favors users with fewer running test runs to prevent anyone monopolizing the
system. With this in place, Fallout now handles over 200 test runs a day. Figure \ref{testruns} shows the mean
number of daily test runs per month. Table \ref{table1} shows additional yearly statistics for this time
period.

%\begin{table}
%\centering
% \begin{tabular}{|c | c | c | c | c | c |} 
% \hline
% Year & Total & Mean & Min & Max\\
% \hline
% 2016 & 759 & 8 & 0 & 44 \\
% 2017 & 5512 & 15 & 0 & 101 \\
% 2018 & 58625 & 160 & 0 & 562 \\
% 2019 & 64633 & 177 & 0 & 361 \\
% 2020 & 62616 & 171 & 0 & 421 \\
% 2021 & 39945 & 197 & 34 & 349 \\
% \hline
%\end{tabular}
%\caption[justification=centering]{Table to test captions and labels}
%\end{table}

\begin{table}
\begin{center}
\begin{threeparttable}
\begin{tabular}{|c|c|c|c|c|} \hline \hline
Year & Total & Mean & Min & Max\\
\hline
2016 & 759 & 8 & 0 & 44 \\
2017 & 5512 & 15 & 0 & 101 \\
2018 & 58625 & 160 & 0 & 562 \\
2019 & 64633 & 177 & 0 & 361 \\
2020 & 62616 & 171 & 0 & 421 \\
2021 & 39945 & 197 & 34 & 349 \\
\hline
%Source & Intensity ($W/cm^2$) \\ \hline
%Experiment & 1.00 \\
%Tan-slab & 1.00\tnote{*} \\
%CHYRA & 1.00\tnote{*} \\ \hline \hline
\end{tabular}
\caption{Test run statistics}\label{table1}
\end{threeparttable}
\end{center}
\end{table}

\subsection{REST API}

The Fallout command-line client is built using a Python library for accessing the Fallout REST API.
Making this API available instead of only providing access to Fallout via the web UI has helped many
other services leverage Fallout's test running capabilities and has no doubt led to Fallout's rise
in popularity at DataStax. Recently, we have used Fallout's API and Python library to drive Fallout
tests using \emph{pytest} \cite{PYTEST} for a new project.

\section{Results}\label{results}

Once one or more benchmarks have been run on a cluster, we use multiple tools to display benchmark
and OS metrics. Fallout includes a built-in way to display client-side benchmark metrics as part of
the web UI but we usually collect many more metrics for runs such as Apache Cassandra and OS
metrics. We use a central Grafana server, known as the history server, to display all of the
historical metrics that are accumulated during test runs.

\subsection{Performance Reports}

\begin{figure*}
	\includegraphics[width=\textwidth]{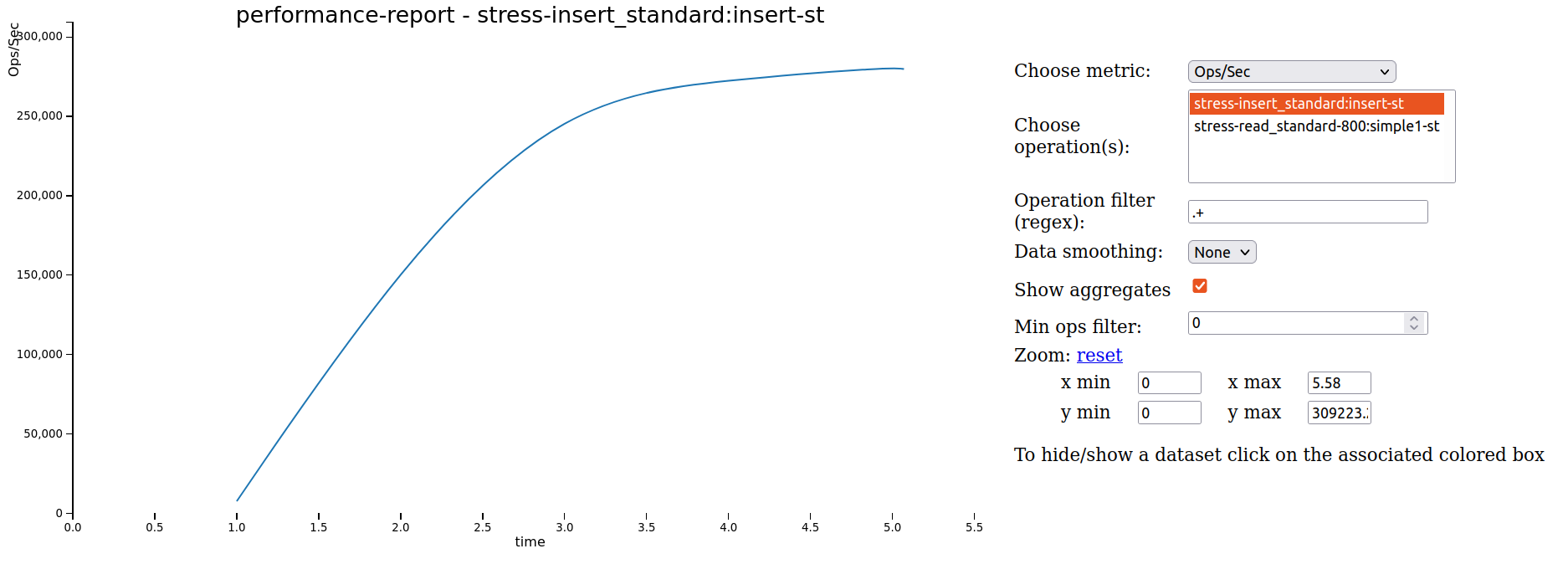}
	\caption{Example performance report}\label{report-single}
\end{figure*}

Fallout can generate performance reports which visualizes the metrics gathered from a single test
run. Performance reports are built on top of HdrHistogram datasets \cite{HDRHISTOGRAM}. The
HdrHistogram format is a de facto standard for histogram data and implementations are available for
many benchmarking tools. A feature that we use heavily is the ability to merge HdrHistogram data
across multiple clients which makes it possible to split load across nodes, collect individual HDR
files, and combine them to summarize the total load on the cluster. Finally, HDR files
capture both throughput and latency in a single file format. Figure \ref{report-single} shows an
example of a performance report.

Metrics are displayed using time series data which is invaluable for database workloads where the
workload does not have a consistent behavior, e.g. where it changes as memory-resident data
structures fill up and are flushed to disk. Being able to see metrics for the entire test run
duration makes it easier for users to spot situations where the test hits an unexpected state. The
metric data used to create graphs can be altered by selecting an item from the drop-down menu on the
right of the page and in this example in Figure \ref{report-single} each phase of the test run
records a separate set of metrics.  Digesting time-series metrics into a single number is impossible
to do manually, so we also provide summary metrics that list throughput, mean, median, and
percentiles for the test run though these metrics are missing from Figure \ref{report-single} above
due to lack of space.

Performance reports are globally readable for all logged in Fallout users and we have used this
feature to share test runs across teams that were collaborating on investigating performance issues
-- having a single location to refer to for a test run's performance helped everyone to agree what
work needed to be done next.

\begin{figure*}
	\includegraphics[width=\textwidth]{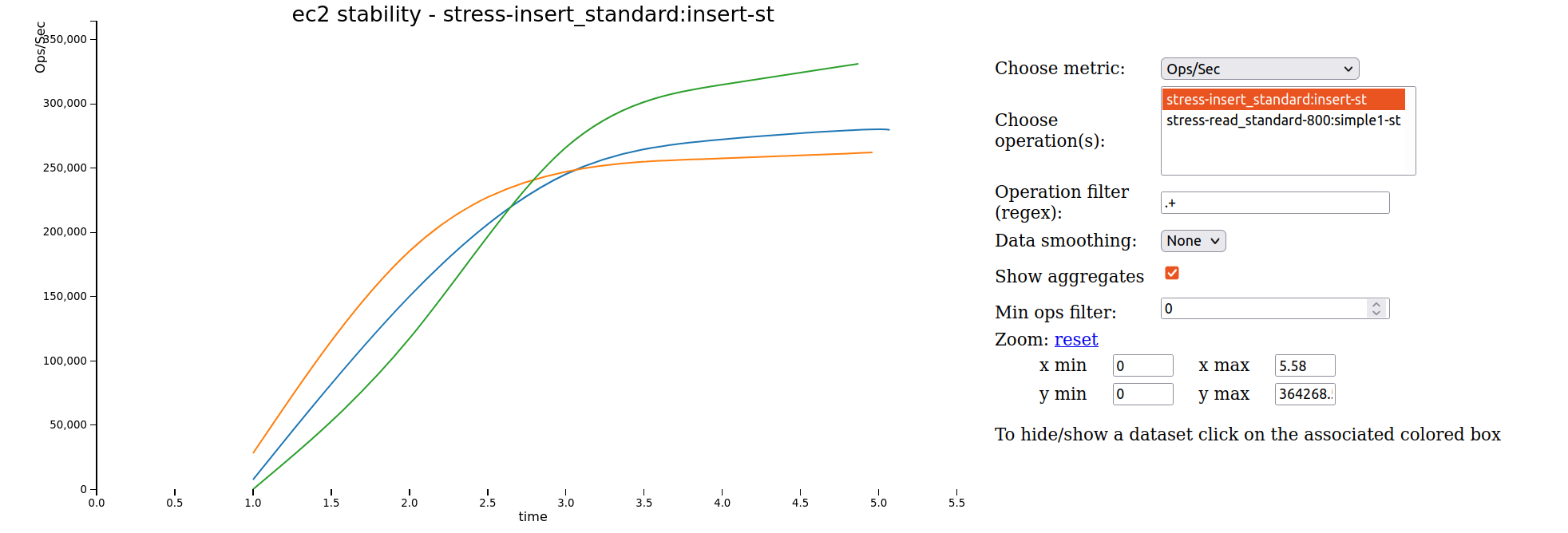}
	\caption{Example grouped performance report}\label{report-multi}
\end{figure*}

Individual performance reports can be grouped together into one report which allows users to look
for differences in performance between test runs. Figure \ref{report-multi} shows an example of a
grouped performance report.

Graphed metrics for each run are displayed in the group report using different colors and details of
the runs are included below the chart in a key which is not included in Figure \ref{report-multi}
again due to lack of space. The group performance reports are particularly useful for comparing different
versions of Apache Cassandra or different configuration options on either the server or client side.
When performance reports started appearing in Jira tickets to illustrate performance
improvements and regressions, we knew that this feature had become successful as a way of quickly
visualizing the performance of benchmarks. Over time, these links to performance reports have become
even more useful as engineers have been able to refer back to previous benchmarking with
ready-to-run tests they can reuse to troubleshoot new issues.

\subsection{History Server}

Though performance reports offer a helpful way to look at the performance of a small number of test
runs for comparison, the fact that all of the metrics from a test run are presented in a time-series
chart makes it unsuitable for analyzing historical trends. When we need to understand how the
performance of our automated tests have changed over the past few days or weeks we use a central
Grafana server we call the history server. This server aggregates OS and application metrics from
both clients and servers for historical analysis and is one of the ways that release engineers
assess the quality of DataStax products. Aggregated metrics are very coarse grained to reduce disk
space usage and calculate simple summary statistics -- each metric is reduced to a single data point
per run regardless of the duration of the test run.

Given that the history server is a central component of quality engineering for releases, it may be
surprising that the hardware resources used to run it are extremely modest. The original version of
the history server ran on a virtual machine with 1 CPU, 4GB of RAM and a 20GB hard disk drive. The
current configuration uses 2 CPUs, 4GB of RAM and an 80GB hard disk drive. We believe that the
reason the history server has survived for many years without any kind of downtime and without
exhausting its small disk space is due to the aggressive graphite retention policy we apply to all
metrics. The default metric namespace, \emph{temporary}, has a retention policy of 1h:15d which
works well for one-off investigations because metrics can be updated once per hour and are
automatically deleted after 15 days. We use a separate namespace, \emph{performance\_regressions},
to retain metrics for much longer but with a reduced frequency: daily metrics are recorded at most
once a day, weekly metrics are recorded once a week, and both are retained for 10 years.
Graphite's design requires that disk space for all configured metrics be allocated up front and
storage for a single metric is 12 bytes, so we can calculate that storing one metric in
\emph{performance\_regressions} every day for a full year only consumes 4.3KB of disk space.

\begin{figure*}
\begin{center}
	\includegraphics[width=\textwidth]{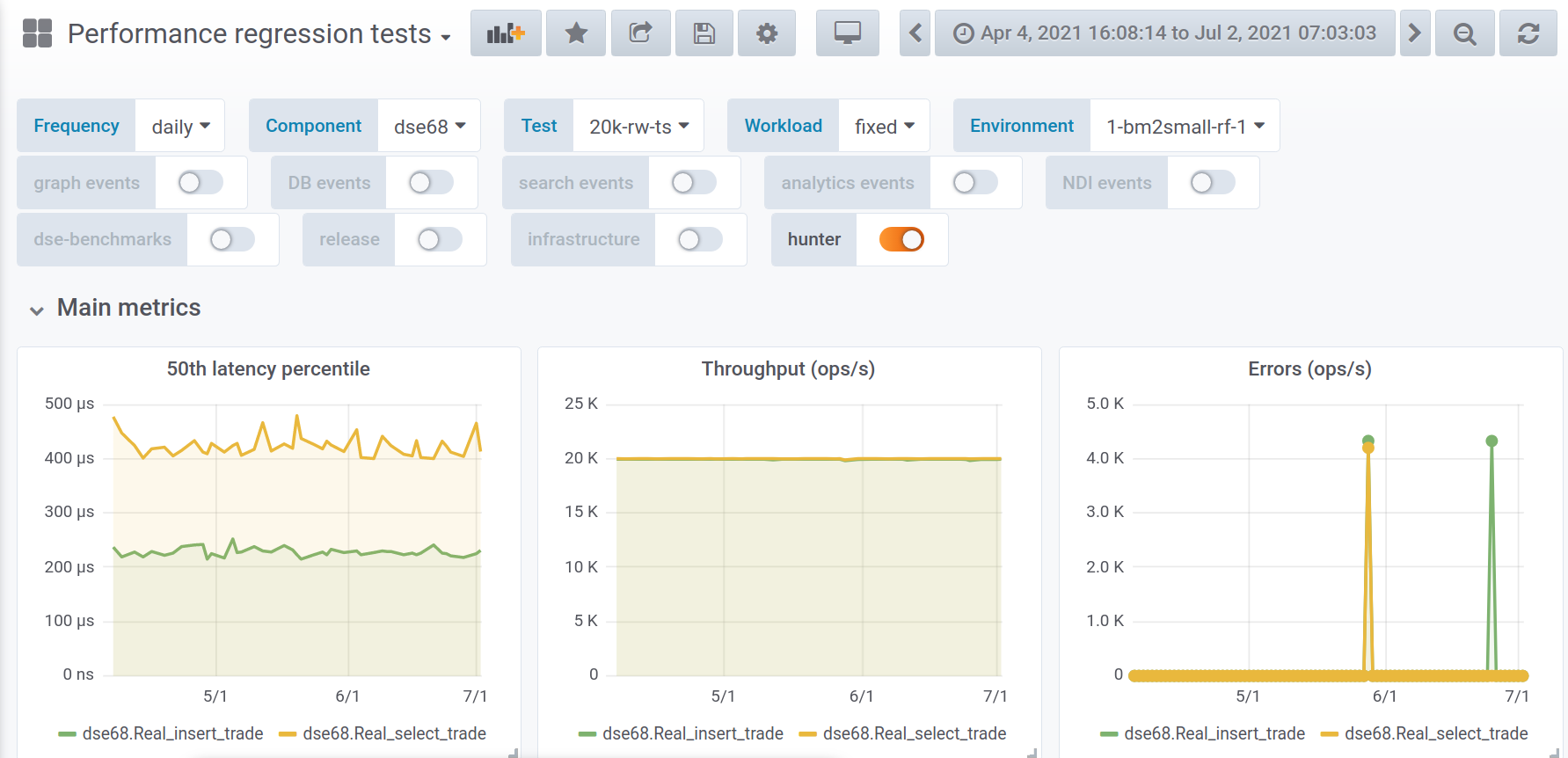}
	\caption{Grafana dashboard}\label{grafana}
\end{center}
\end{figure*}

Figure \ref{grafana} shows one of the Grafana dashboards from the history server which includes panels for
throughput, error count, and percentile metrics.

\section{Lessons Learned}\label{lessons}

Fallout has evolved over many years of development and we have found that while some of our initial
design choices were correct and have stood the test of time, others were wrong and needed
reassessing. And some problems we never even anticipated.

\subsection{Configuration Files Should Be Short and Expressive}

The more lines a test configuration file has the greater the chance of introducing a bug. One of the
goals for Fallout has been to provide enough support in the test and benchmark modules that
common use cases only need small test run configuration files which reduces the probability that a
user will make a mistake. This is still an on-going effort as it takes time for common usages to
emerge when support for new modules is added but the end result is happier users with greater
confidence in Fallout. This goal has served us well in creating a useful configuration language that
is easy to understand.

\subsection{Templating for Configuration Files Encourages Reuse}

As Fallout amassed more users and the number of test run configurations increased, we noticed that
many users began copying and pasting YAML across config files. A common situation where this happens
is when users need to run the same test across multiple versions of an app, e.g. running the same
benchmark against Apache Cassandra 3.11 and 4.0 to compare performance. We added support in
Fallout's YAML parser for mustache \cite{MUSTACHE} templates which allow users to use templates in their
YAML files and provide specific values either on the Fallout test run web page or as parameters via
the REST API.

Even with mustache templating, we found that users wanted to separate out common chunks of YAML into
different, smaller files and include them in multiple configuration files. Additionally, users wanted
to be able to store these files in a version control system. Fallout does not support either of these
features so the haxx project was created which uses Jinja \cite{JINJA} templating to allow composition of
test fragments and to provide version control via a git repository.

\subsection{Tests Need Access to External Files}

A feature that we failed to anticipate early on was that tests, benchmarks, and tools would need
the ability to access external files, e.g. configuration files. We initially worked around this
limitation by either extending the test module to fetch the external file from a GitHub gist or by
generating the test config file at runtime based on the keys and values in the Fallout YAML config.
This approach did not scale as we added new modules and it is now possible to use a unified method
to access external files with the \emph{<<file:filename>>} syntax regardless of the module used in
the test run config.

\subsection{Long-running Tests Benefit From Semantic Checks and Idempotency}

It is very straightforward to check YAML files for syntactic errors and there are numerous Java
libraries available to do that, such as SnakeYaml \cite{SNAKE} which is the library that Fallout uses.
However, syntactic errors are only one source of problems afflicting users. Since most of the YAML
values in a test config are consumed by tools other than Fallout, it is challenging to validate that
the semantics of those values behave as expected. We have encountered situations where a single
mistyped character in a NoSQL table name caused all subsequent test phases to fail and was only
triggered after the test had been running for an hour.

Additionally, re-running Fallout tests sometimes requires the infrastructure to be torn down and
brought back up if Fallout cannot determine the runlevel of the cluster. Other deployment tools,
such as Terraform \cite{TERRAFORM} solve this problem with idempotency which allows the same deployment
steps to be applied repeatedly without causing any changes to the underlying machine if the
corresponding configuration for those steps has not changed. Fallout does make an attempt to detect
the current cluster runlevel and skip unnecessary configuration steps but the detection is
imperfect. This detection is used in Fallout's cluster-reuse mechanism, which is triggered by naming
a cluster and requesting that it be left in a specific runlevel at the end of a test run; subsequent
test runs with the same test definition will find the named cluster, detect its runlevel, and
continue from there. This makes it possible to iterate on test creation a little bit faster, and—in
some specialized cases -- skip slow data loading steps for big-data tests. However, in our experience
most users do not encounter situations where they need to use these features.

\section{Related Work}\label{related}

Automated testing, which includes running benchmarks, is a vital part of ensuring quality for
software projects \cite{MOOSHOT}. Integrating benchmarking into a continuous deployment pipeline is
discussed in \cite{CB} which focuses on using performance metrics with thresholds to decide whether
changes should be allowed into production. Since we use Fallout to test software that will
ultimately be deployed to a variety of environments, ranging from the cloud to on-premise, there is
no built-in functionality for gating deployments based on performance change thresholds. Instead,
statistically significant changes are detected using change point detection and a developer is
required to make the deployment decision. Automating deployments with Fallout is one of our future
goals.

MockFog 2.0 \cite{MOCKFOG} enables fog applications experiments by emulating fog infrastructure in
the cloud and has a very similar design to Fallout. Both MockFog and Fallout  provision
infrastructure, configure and deploy applications, run tests and benchmarks, and even use states
(Action states and NodeGroup states, respectively) to define legal transitions for the internal
state machine. However, MockFog uses Docker to manage applications whereas Fallout supports both
native and Kubernetes-based applications which more closely aligns with typical deployments of
Apache Cassandra and Apache Pulsar. MockFog also uses Ansible to configure infrastructure which
provides the idempotent state updates that are partly missing from Fallout's implementation.

Adelphi \cite{ADELPHI} is an open-source QA tool that runs on top of Kubernetes and allows users to run
data integrity and performance tests against Apache Cassandra. It is packaged as a helm chart and
includes a limited number of benchmarks and testing tools so that users can compare two clusters
against one another. Adelphi takes care of executing the tests but does not provide facilities to
create and terminate the underlying Kubernetes clusters or present the benchmark and test results
for analysis.

MongoDB's Distributed Systems Infrastructure (DSI) \cite{DSI} was developed at approximately the same
time as Fallout though the two projects were not known to each other. DSI shares many things in
common with Fallout including components to provision virtual machines, configure database servers
and benchmarks, collect results for automated and visual inspection, and finally teardown the
infrastructure when the test completes. Both Fallout and DSI use YAML configuration files to control
test runs. However, Fallout differs from DSI in a number of ways. Fallout is written in Java and
DSI is written in Python. While DSI primarily targets Amazon EC2, Fallout can currently launch tests
on Google Cloud Platform, Amazon EC2, Microsoft Azure, as well as our internal OpenStack-based
private cloud. Because ctool already existed when Fallout was created, Fallout has a very modular
architecture and relies on other tools and components to do certain tasks whereas most of the
corresponding functionality for DSI is built into the service. Lastly, as far as the authors are
aware, DSI does not expose an API for other tools to call.

Work on reducing the cost of testing very large distributed systems by running many virtual machines
on top of fewer physical servers is discussed in \cite{DIECAST}. This work targets network services with
thousands of nodes which are much larger than typical Apache Cassandra or Pulsar clusters.

RocksDB includes tools for running benchmarks and analyzing the results but no project exists to
handle the setting up and tearing down of hardware to run the benchmarks \cite{ROCKSDB}. Likewise, SAP
has published work that shows how they integrate performance testing into their CI process \cite{SAP} but
no details are included on the way that tests are deployed on their testing infrastructure.

\section{Conclusion}\label{conclusion}

Fallout is a distributed systems testing service capable of automatically provisioning clients and
servers, installing, configuring and executing distributed apps and workloads, and centrally
collecting results for later analysis. We use Fallout internally at DataStax and it drives the
entire performance and testing ecosystem for both our Apache Cassandra and Apache Pulsar products.
Fallout started life with a very specific purpose and has evolved after years of engineering effort
to be the backbone of performance and quality for us and it provides our engineering teams with
fully-automated end-to-end testing for distributed systems. Fallout's REST API has been essential
for new teams to leverage Fallout's distributed testing and has encouraged the birth of numerous
tools and services that complement Fallout. Our Fallout server executes around 200 tests every day,
and on busy days runs closer to 400 tests.

Since each of our engineering teams have their own preferences for the kinds of benchmarks, cluster
configurations, and cloud infrastructure, all of these components are configurable in Fallout which
has been designed with modularity in mind. We have extended this modularity to allow tests and
benchmarks to load external files and added templating so that users can reuse test config fragments
without copying and pasting.

We have released Fallout as an open-source project with the hope that the open-source community can
benefit from our investment and the lessons we have learned running Fallout in production for over 5
years.

\section{Acknowledgements}\label{acknowledgements}

We would like to thank the anonymous reviewers for their valuable feedback and suggestions. Fallout
was created by Jake Luciani, Joel Knighton, and Philip Thompson and we are thankful for their
decision to create a new tool to solve the complex problem that is distributed systems testing. The
history server was created by Pierre Laporte and it is stability is illustrated by the fact that it
has been the component that has required the fewest updates in the whole Fallout ecosystem.
Christopher Lambert was largely responsible for integrating ctool support to Fallout and James
Trevino continues to maintain and improve Fallout. Ulises Cerviño Beresi created haxx. Shaunak Das
contributed numerous test modules. Many more people have contributed to Fallout and related testing
services and we are grateful for all of their efforts.

The open-source version of Fallout owes a great deal of gratitude to Jake Luciani and Jonathan Ellis
for championing the project internally at DataStax.

%% Loading bibliography style file
%\bibliographystyle{model1-num-names}
\bibliographystyle{cas-model2-names}

% Loading bibliography database
\bibliography{bench21-fallout}

% Biography
%\bio{}
% Here goes the biography details.
%\endbio

%\bio{pic1}
% Here goes the biography details.
%\endbio

\end{document}